\documentclass[twocolumn,preprintnumbers,
showpacs,amsmath,amssymb,aps,prd,nofootinbib]{revtex4-1}
\usepackage{graphicx}
\usepackage{color}
\usepackage[utf8]{inputenc}
\usepackage[colorlinks,hyperindex]{hyperref}  % Si on compile en PdfLaTeX
\usepackage{accents}
\usepackage{enumitem}

\hypersetup
 {
   colorlinks,%
   citecolor=blue,%
   linkcolor=blue,%
   urlcolor=blue,%
 }
\def\setR{\mathbb{R}}

\def\ie {i.e.,~}

\newcommand{\sss}[1]{\scriptscriptstyle #1} 

\newcommand{\pbundle}[4]{#1(#2,#3,#4)}%Notation cf KobayashiNomizu
\newcommand{\sbundle}[3]{#1(#2,#3)}

\newcommand{\Mov}[1]{{\color{magenta}{#1}}}
\begin{document}
%\preprint{APS/123-QED}

\title[TEGR as gauge theory:comments and issues]{Teleparallel theory as a gauge theory of translations~: comments and issues.}

\author{M.~Le~Delliou$^{1,2}$, E.~Huguet$^3$, and M.~Fontanini$^3$}
\affiliation{$1$ - Institute of Theoretical Physics, School of Physical Science and Technology, Lanzhou University,
No.222, South Tianshui Road, Lanzhou, Gansu 730000, P R China
} 
\affiliation{$2$ - Instituto de Astrof\'isica e Ci\^encias do Espa\c co, Universidade de Lisboa,
Faculdade de Ci\^encias, Ed.~C8, Campo Grande, 1769-016 Lisboa, Portugal}
\email{(delliou@lzu.edu.cn,)morgan.ledelliou.ift@gmail.com}
\affiliation{$3$ - Universit\'e Paris Diderot-Paris 7, APC-Astroparticule et Cosmologie (UMR-CNRS 7164), 
Batiment Condorcet, 10 rue Alice Domon et L\'eonie Duquet, F-75205 Paris Cedex 13, France.}%\\   
\email{michele.fontanini@gmail.com\\
huguet@apc.univ-paris7.fr}%\\
\date{\today}% It is always \today, today,
%\pacs{}% PACS, the Physics and Astronomy
%Use showkeys class option if keyword
\begin{abstract}
%TEGR 
The Teleparallel Equivalent to General Relativity (TEGR) is often presented as a gauge theory of 
translations, \ie
that
uses \emph{only} the 
translation group $T_4 = (\setR^4, +)$
as its gauge group. In a previous work 
we argued 
against this \emph{translation-only} 
formalism on the basis of 
its mathematical 
shortcomings. We then provided 
an alternative proposal using a Cartan connection. 
Recently, a reply by some of the authors defending TEGR as a \emph{translation-only} gauge theory discussed our objections.
Here, we first clarify our 
arguments, and give new proofs of some statements, to answer to these discussions, maintaining our first
claim.  We then 
amend one of the argument that originally  
led us to propose
the Cartan connection in this context. This 
broadens the \emph{a priori} possible choices for a TEGR connection.
\end{abstract}
\maketitle
\tableofcontents
\section{Introduction}

The Teleparallel Equivalent to General Relativity (TEGR) theory is an alternative formulation for the classical theory of gravity, equivalent
to General Relativity (GR) - both yield the same predictions. The TEGR gravitational field is carried by the torsion tensor, the curvature being zero. Consequently, the two theories, GR and TEGR, yield 
a completely different interpretation 
of the effects of gravitation. The interest for TEGR has increased 
in the past two decades, among other reasons because it serves as a starting point 
for other proposal for gravity, such as\footnote{See references and comprehensive summaries of TEGR therein.} $f(T)$ \cite{Ferraro:2006jd} and $f(R, T)$ \cite{Harko:2011kv, Bahamonde:2015zma} theories, Symmetric TEGR (STGR) \cite{Nester:1998mp, BeltranJimenez:2017tkd}, 
Conformal TEGR \cite{Formiga:2019frd, Bamba:2013jqa}.  
Another major point of interest for 
TEGR is its formulation as a classical gauge theory (see \cite{Blagojevic:2013xpa} for comprehensive review of 
gauge theory of gravitation): TEGR is often presented as 
a gauge theory for the (four dimensional) 
translations group, hereafter $T_4 \equiv (\setR^4,+)$. Such formulation 
would reconcile Gravity with the other known interactions (electromagnetic, weak and
strong), since it would 
associate
gravity, at classical level, to a gauge field. 

In a recent paper \cite{Fontanini:2018krt}, we 
argue against this last description in which TEGR is a gauge theory of {\it only translations}, and suggest an alternative possibility (yet to be fully explored) to recover a gauge theory formulation for TEGR. The central concern with the translation-only gauge approach comes from its incomplete 
fit to 
the admitted mathematical description of gauge theories, despite taking into account 
the peculiarities (in particular the soldering property) of TEGR with respect to 
the other (particles physics) gauge theories. 

In turn, these arguments have 
recently been criticized by Perreira and Obukhov \cite{Pereira:2019woq}, in a paper 
which elaborates on the 
mathematical structure of the theory (in particular that of the principal translation bundle). 

Furthemore, 
one of the conclusions of \cite{Fontanini:2018krt} lead 
both translations and local Lorentz transformations 
to compose 
the connection form (gauge field) in order to obtain 
TEGR. A simple 
connection
proposal was formulated, for which the precise form of the translation term, apart from its abelian character, was not specified. 
Consequently, the curvature for 
this connection appeared to contain 
a cross-term, contrary to the curvature for  
TEGR (namely the torsion). However, it turns out that
the form of the translation term of  
the connection is prescribed,  
with the consequence 
that 
the cross-term always vanishes. This doesn't changes the conclusions drawn in \cite{Fontanini:2018krt}, but opens the possibility for 
more connections types than were proposed in \cite{Fontanini:2018krt}.

In the present paper, we first take the opportunity of the arguments given and criticisms risen in  \cite{Pereira:2019woq} to 
sharpen the main arguments of \cite{Fontanini:2018krt} and give 
new proofs of some statements. We also take into account
the form of the translation term in the Ehresmann connection discussed 
in \cite{Fontanini:2018krt}, give its 
generic curvature and briefly discuss the other possibilities besides 
the proposed Cartan connection. 

For definitions not explicitly stated, we refer to \cite{Fecko:2006,Nakahara:2003,Isham:1999,KobayashiNomizu:1963}.
Throughout the paper we denote by $\pbundle{E}{M}{F}{\pi}$ a fiber bundle with total space $E$, typical fiber $F$, four 
dimensional differentiable base manifold $M$ and projection $\pi$, and we sometimes shorten this notation to 
the total space $E$ 
when no confusion is possible.

\section{What is meant by gauge theory
}\label{SEC-1-Terminology}
Although the notion of gauge invariance was introduced by H. Weyl in the context of conformal (Weyl) spaces, and in an attempt to generalize the 
General Relativity, its current meaning is 
rather broad and related to  
the solutions of some equations, 
determined up to 
a symmetry. In the 
paradigmatic example, 
the Maxwell equations for the electromagnetic potential, solutions are determined 
up to 
a derivative. Gauge 
theory, coined in  
particle physics, describes a more restricted physical situation:  
a Lagrangian for some 
field $\psi$ (usually a matter field), invariant under a global (\ie 
spacetime-independent)  symmetry,  gets 
modified to become invariant 
under the local (\ie 
spacetime-dependent) 
version of the symmetry, by 
coupling the field $\psi$ to
an additional field $A$, so that the modified 
Lagrangian is 
local-symmetry 
invariant. In practice, 
this procedure amounts to replace  
by covariant derivatives the original (matter field) Lagrangian derivatives, where 
the additional field $A$ appears as a connection. The new field's dynamical term is 
generated by the so-called field-strength $F$\Mov{, }
formally the curvature of %
the connection (field) $A$.  As  
the free field $A$ equations 
exhibit gauge invariance, 
$A$ is termed a gauge field.   

Gauge theories 
of classical fields are well described mathematically. In particular, the gauge field is represented
by a connection (of Ehresmann type) in a principal 
fiber bundle -- \ie{with 
structure group and typical fiber given by the group of symmetry of the theory} --  
and its field-strength corresponds to that connection's curvature. 
The following assumes this geometrical framework as the correct description for gauge fields and thus that the corresponding physical quantities and structures should be properly related in order to speak of gauge theory.

\section{Issues 
with a gauge theory of translations}\label{SEC-2-Issues}
In a previous paper \cite{Fontanini:2018krt} we show that the usual description of TEGR as a gauge theory of \emph{translations} is not mathematically well defined. 
The 
main arguments we present are 
that~:
\begin{enumerate}
\item \label{enu:TranslationConnection}the connection in the principal bundle of translations, not including any relation to the Lorentz group, cannot yield the TEGR curvature, which is the torsion. In particular this is because the  
torsion is defined through the canonical form, 
itself defined in the bundle of frames,
\item \label{enu:trivialT4}the principal bundle of translations is trivial (a product space), and thus 
inappropriate to describe TEGR 
as  equivalent to General Relativity,
\item \label{enu:T4notTangent}the principal bundle of translations does not identify with the tangent bundle, contrary to what is stated in translation-only gauge theory \cite[described for instance in][]{Aldrovandi:2013wha}.
\end{enumerate}
Recently, Pereira and Obukhov \cite{Pereira:2019woq} have criticized part of these arguments in a more detailed version of their account of the gauge structure of TEGR. We  
take the opportunity of these criticisms 
to append some details 
to our arguments, add some new proofs, and  clarify our aims.

We start with 
the gauge theoretic bundle framework of TEGR (points \ref{enu:trivialT4} and \ref{enu:T4notTangent}  
above). We point out 
that, contrary to related in \cite[second paragraph of sec. 3]{Pereira:2019woq}, our argument in \cite{Fontanini:2018krt} does 
not assume that the principal bundle of TEGR as a gauge theory is the frame bundle. In fact, we mention the frame bundle as a 
part of our conclusions: the ``non-standard'' gauge translation view of TEGR \cite[as described in][]{Aldrovandi:2013wha} is, in our opinion, not a gauge theory of translations alone, but a (correct) formulation of TEGR in the frame bundle for 
which the translational part is the canonical form, 
(mistakenly) interpreted as a gauge potential.

\subsection{The 
principal bundle of translation and its triviality
}\label{SUBSEC-21-PrinTransBund}
Following
the standard geometrical description of gauge theories, to represent 
the gauge theory of translations, we have to use 
the principal translation bundle $\pbundle{P}{M}{T_4}{\pi}$, where $T_4$ is the four dimensional group of translations
 -- \ie $(\setR^4, +)$  
-- $M$, the base (spacetime) manifold 
and $\pi$, the projection onto the base.
As is the case for all principal $G$-bundle (\ie
with structure group and fiber $G$), the typical fiber of $\pbundle{P}{M}{T_4}{\pi}$ is an homogeneous 
space (synonymous with 
affine $G$-space, or $G$-torsor). For such space 
the action of the group $G$ is defined as 
\begin{enumerate}%[1)]
\item free (\ie no group element \label{enu:free}
except the identity --the neutral element $e$ -- leaves any point fixed)
 and
\item transitive (\ie any two points can be related through the action of $G$).
\end{enumerate}
In particular, each 
fiber is the single
orbit under
the action of $G$. Note that the Minkowskian scalar product is not needed in the definition of $\pbundle{P}{M}{T_4}{\pi}$.  
Furthermore, we stress out that the translation bundle $\pbundle{P}{M}{T_4}{\pi}$ structure group is $(\setR^4, +)$, in contrast with the tangent bundle $TM$ structure group GL$(4,\setR)$, so 
$\pbundle{P}{M}{T_4}{\pi}$ and $TM$ cannot
be identified as bundles.  

We prove 
in \cite{Fontanini:2018krt} that the principal bundle of translations is trivial (\ie 
a product space $M\times\setR^4$), or more exactly 
that its base spacetime manifold can only be  
associated with a trivial frame bundle, thus restricting too much the type of spacetimes the resulting gauge theory can produce, 
compared to GR.
There is, however, 
another, more direct proof of this statement, based on  
the notion of classifying space. Interested readers can find the details of proof and references on classifying spaces 
in App.~\ref{App-A}. To summarize the argument: 
if the classifying space of a principal bundle reduces to a point then the bundle is trivial. 
This is the case 
for the bundle of translations whose classifying space is the quotient of $\setR^4$, which is contractible to a point, by the translations. 

We finally point out that the statement \cite[][end of sec. 3]{Pereira:2019woq}  
`` \emph{\ldots the bundle of teleparallel 
gravity is not a vector bundle. Consequently, it does not admit a global section, and is in general nontrivial.}'' is not correct as formulated: 
that a bundle is not a vector bundle does not guarantee that it cannot admit a global section. For instance the 
principal bundle $(M\times G)(M,G,\pi)$, which is trivial by construction, admits a global section $s_g$ defined as 
\begin{align*}
 s_g:~ & M\mapsto P \\
	& x\longrightarrow s_g (x):= \phi(x,g),
\end{align*}
where $g$ is a fixed element of $G$ and $\phi$ a trivialization.

\subsection{The connection for translation and its curvature}\label{SUBSEC-22-TransGaugeFieldAndCurvature}
Let us now turn to the point \ref{enu:TranslationConnection} of our arguments against the interpretation of TEGR as a gauge theory 
of {\it only} translations: namely 
the status of the connection for the translation 
symmetry. The point of view of the translation 
gauge theory is well summarized in 
\cite[][sec. 1.2]{Pereira:2019woq}. Recall the pulled back connection onto 
the base manifold along a partial 
section $\sigma$ \cite[Eq. (7) of][]{Pereira:2019woq}: 
\begin{equation}\label{EQ-ConnectionOnBase}
 h^a:=\sigma^* \omega_{\sss T}^a,
\end{equation}
where $\omega_{\sss T}^a$ are the components in the Lie algebra of the translation group,  
Lie$(T_4)$, of the connection  $\omega_{\sss T}$.
The usual formula ($\Omega = d\omega + \omega\wedge\omega$) can be used to calculate the curvature $\Omega_{\sss T}$ of this connection,
which in this Abelian case (translations) reduces to $\Omega_{\sss T} = d\omega_{\sss T}$. 

In the usual presentation of the translation gauge view of TEGR, $h^a$ is said identified with a coframe, and $\Omega_{\sss T}$ with the torsion for that coframe  
\cite[sec. 1.2]{Pereira:2019woq}. These 
identifications are raising an issue, since 
torsion (and its pullback along some section $\sigma$) 
are usually defined through the so-called canonical form $\theta$. Precisely, that form is defined \emph{on the frame bundle:} 
$\pbundle{LM}{M}{Gl(4,\setR)}{\pi}$, 
or its restriction to orthonormal frames $\pbundle{LM}{M}{SO_0(1,3)}{\pi}$, through the expression:
\begin{equation*}
 (\theta(e), \xi) = (e^{-1}, \pi_* \xi),
\end{equation*}
or in component
\begin{equation*}
 (\theta^a(e), \xi) = (e^a, \pi_* \xi),
\end{equation*}
where $e$ is a frame in $LM$  over a point $x$ of the base manifold $M$, $\xi$ a vector of LM, and $\pi$ the projection on the base.
Along a section $\sigma$ of the frame bundle one has:
\begin{equation}\label{EQ-FrameEqSecOfCanonicaForm}
 e^a =\sigma^* \theta^a.
\end{equation}
This expression looks very similar to (\ref{EQ-ConnectionOnBase}) and one is tempted to identify $\theta$ with the connection for translations $\omega_{\sss T}$, but these two mathematical objects are very different since they are defined in two distinct (and non-isomorphic) bundles: 
the connection $\omega_{\sss T}$ is an $\setR^4$ -valued  (Lie$(T_4) = \setR^4$) one-form  on $P$, whereas $\theta$ is an $\setR^4$-valued one-form 
on $LM$. In addition, we note that nothing forbids the term $h^a$ of Eq. \ref{EQ-ConnectionOnBase} to vanish, which is not allowed for a (co)frame.

This identification problem between $\omega_{\sss T}$ and $\theta$ relates  
to the problem of identification between $\pbundle{P}{M}{T_4}{\pi}$ and $TM$: the canonical form $\theta$, as the pull-back on $LM$ of the soldering form $\widetilde\theta$ on $M$ (see \cite{Fontanini:2018krt} for definition and references), realizes the so-called soldering. In the present case of the frame bundle soldering, $\widetilde\theta$ is the identity map between $TM$ as the tangent bundle of $M$, and $TM$ viewed as an associated vector bundle of $LM$.  It is therefore loosely consistent, while   
identifying $TM$ to $\pbundle{P}{M}{T_4}{\pi}$, 
to identify $\omega_{\sss T}$ and $\theta$ through the expressions (\ref{EQ-ConnectionOnBase}) and  (\ref{EQ-FrameEqSecOfCanonicaForm}).  
These identifications are unfortunately not allowed from the mathematical point of view. 
  
\subsection{On 
the gauging of the Lorentz group}\label{SUBSEC-23-CommentGaugingLorentz}
In Sec.~\ref{SUBSEC-22-TransGaugeFieldAndCurvature} was already pointed out  
that gauging the translations only leads to the 
curvature $\Omega_{\sss T} = d\omega_{\sss T}$. Leaving aside the identification between the canonical form $\theta$ and the translation connection $\omega_{\sss T}$ discussed in the paragraph \ref{SUBSEC-22-TransGaugeFieldAndCurvature}, the full torsion $d\theta + \omega \wedge \theta$ requires the addition, in the curvature of the TEGR bundle, of the second term $\omega \wedge \theta$, where $\omega$ is the Lorentz (or spin) connection.

This addition is claimed, in 
the usual translation gauge 
interpretation of TEGR as summarized in \cite{Pereira:2019woq}, to take 
into account non-holonomic frames while stating it does not 
relate to the gauging of
the Lorentz symmetry.
In particular, since there is no dynamics associated with the Lorentz term $\omega$ of TEGR  (the Weitzenb\"ock connection, its corresponding strength field, the curvature, is zero) 
it thus would not be able to 
be a gauge field. In our 
view, this addition 
reads as the replacement of the exterior derivative $d$ by a covariant version $d + \omega \wedge$, thus on mathematical grounds points toward 
a gauging of the Lorentz symmetry, $\omega$ defining an (Ehresmann) connection in $LM$. 
Indeed, the two point of view rely on different aspects of gauge theory. On the one hand it is true that 
the Lorentz connection alone, set to the Weitsenb\"ock connection, is not  a gauge field \emph{per se}, in the sense that it does not mediate the interaction. On the other hand it is also true that the Lorentz connection is
introduced to enforces the local Lorentz invariance of the theory. Thus, these two statements do not  contradict each other. In addition, since, from our point of view, this discussion leaves aside the
identification problem between the translation field $\omega_{\sss T}$ 
and the canonical form $\theta$, it 
cannot alone provide  
an argument for or against the translation gauge approach of TEGR.

Nevertheless,
and independently of  
the interpretation one chooses, in order to obtain the TEGR connection curvature to yield the torsion, both symmetry groups,
translation and Lorentz, are required to provide corresponding terms in the connection.

\section{Possible forms of  
connection for TEGR.}
The previous  
observation leading to the need for both translations and Lorentz terms motivates us in \cite{Fontanini:2018krt} to first introduces an ansatz for the Ehresmann connection of TEGR. Our "naive" guess 
sets the total connection as the sum of  
contributions 
for each symmetry: $\omega := \omega_{\sss L} + \theta_{\sss T}$.
The resulting curvature reads 
\begin{align*}
\Omega := &\,d\omega + \omega\wedge\omega\\
=&\,d\omega_{\sss L} + \omega_{\sss L}\wedge\omega_{\sss L} + d\theta_{\sss T} + 
\omega_{\sss L}\wedge\theta_{\sss T} + 
\theta_{\sss T}\wedge\omega_{\sss L}\\
=&\, \Omega_{\sss L} + \Theta_{\sss L} + 
\theta_{\sss T}\wedge\omega_{\sss L},
\end{align*}
where $\Omega_{\sss L}$ and $\Theta_{\sss L}$  are the curvature and
torsion associated to $\omega_{\sss L}$ and $\theta_{\sss T}$. The 
term $\theta_{\sss T} \wedge \theta_{\sss T}$ vanishes due to the abelian character of 
translations.

However, we ignore in \cite{Fontanini:2018krt} that
the form of the translation part 
$\theta_{\sss T}$ is prescribed. As a consequence of that specific form, the cross term always vanishes. In the five dimensional matrix representation where
\begin{equation*}
\theta \mapsto
\begin{pmatrix}
  0 &\theta_{\sss T}\\0 & 0
  \end{pmatrix},~~~~\omega_{\sss T} \mapsto
  \begin{pmatrix}
  \omega_{\sss T}& 0\\0 & 0
  \end{pmatrix},
\end{equation*}one has
\begin{equation*}
\begin{pmatrix}
  0 &\theta_{\sss T}\\0 & 0
  \end{pmatrix} 
  \wedge
  \begin{pmatrix}
  \omega_{\sss T}& 0\\0 & 0
  \end{pmatrix} 
  = 0.
\end{equation*}

This
vanishing of this cross term cannot therefore stand as 
a criterion to choose the
connection, and one has to amend our claims about the composite
Poincar\'e connection  discussed in  \cite{Tresguerres:2012nu}: at least on the base manifold its curvature is the sum
of the Lorentz 
curvature and of the torsion. It thus restricts to the torsion, as needed to describe TEGR, when $\omega_{\sss L} = \omega_{\sss W}$, the Weitzenb\"ok connection.

Note however, that our proposal for the use of a Cartan connection is not affected by the above considerations. Indeed it remains a possible choice in which the soldering property is naturally taken into account in the definition of the form itself.

\section{Conclusion}\label{SEC-Conclusion}
We have clarified  
our arguments about our claims that  
TEGR cannot be considered as a gauge theory for 
translations only. To summarize roughly our main point (see the main text for details): 
\begin{itemize}
\item torsion, \ie the curvature of the connection in a gauge view of TEGR, is built from 
the canonical form $\theta$ and the Lorentz connection $\omega$, that are both defined on the frame bundle. In the translation-only gauge formalism summarized by Pereira an Obukhov in \cite{Pereira:2019woq}, the canonical form is identified with the translation connection $\omega_{\sss T}$, a one-form defined on the bundle of translations-only $\pbundle{P}{M}{T_4}{\pi}$. This identification is not mathematically allowed.
\end{itemize}
Note that %the fact that
the canonical form realizes soldering and %thus
relates the frame bundle to the tangent
bundle $TM$ (see \cite{Fontanini:2018krt} for details). 
%does not
This does not change our main point  since the identification between $TM$ and the translations-only bundle $P$ 
made in translation gauge formalism \cite{Pereira:2019woq} is not allowed either.

In addition to  
these clarifications, we have amended the ansatz made in \cite{Fontanini:2018krt}, originally leading 
us to propose the Cartan connection  
for TEGR. Taking  
into account the specific form of the translation term
in the curvature calculation, all connections of 
the
form $\omega = \omega_{\sss L} + \theta_{\sss T}$, where $\omega_{\sss L}$ is a Lorentz connection and
$\theta_{\sss T}$ stands for the translation part, are \emph{a priori} allowed. 

Finally, independently of the choice of 
connection, the ultimate criterion for selecting a gravitational gauge field proceeds from its coupling 
to matter, which has to be consistent with the observationally tested Levi-Civita coupling. Further investigations should be done on that subject.

\section*{Acknowledgements}

The authors wish to thanks, D. Bennequin for helpful discussions on geometry. The work of M.~Le~D. has been supported by 
Lanzhou University starting fund, and the Fundamental Research Funds for the
Central Universities (Grant No.lzujbky-2019-25).

\appendix
\section{Classifying spaces and triviality of the translation bundle}\label{App-A}

The notion of classifying space comes from homotopy theory and is used in algebraic topology (see for instance \cite{Nakahara:2003} for an 
introduction). In this short appendix, we use it to show the triviality of the translation bundle. For details on definitions and proofs 
of the properties used, the Reader is referred to introductory lectures notes on classifying spaces of \href{https://ckottke.ncf.edu/docs/bundles.pdf}{Kottle} \cite{Kottke:2012} 
and \href{http://www.math.washington.edu/~mitchell/Notes/prin.pdf}{Mitchell} \cite{Mitchell:2011}, as well as to the short presentation in Isham \cite[sec. 5.1.8 ]{Isham:1999}, and to the more advanced treatment on algebraic topology of May \cite{May:1999}. 
Here we use the notation $\sbundle{E}{B}{F}$ for a bundle of total space $E$, base $B$ and fiber $F$ since 
the projection needs not to be specified.

Firstly, let us consider the notions of classifying space and universal bundle.
For a Lie group $G$ one can always find a contractible space, usually denoted $EG$, on which $G$ acts freely.
The classifying space $BG$ of $G$ is the quotient space of $EG$ by the action of $G$. It can be shown that $BG$ is unique for a given $G$, and that the bundle $\sbundle{EG}{BG}{G}$ is a principal $G$-bundle, called a universal bundle. This last name comes from the following property: 
for any principal $G$-bundle $\sbundle{E}{B}{G}$ -~whose base $B$ is a CW space or is paracompact, which is always realized for 
a differentiable manifold~- one can find an isomorphism $f: B \mapsto BG$, up to an homotopy, such that the bundle 
$\sbundle{EG}{BG}{G}$ is the pullback bundle of $\sbundle{E}{B}{G}$, while that pullback is an isomorphism between 
$\sbundle{E}{B}{G}$ and $\sbundle{EG}{BG}{G}$. 

Secondly, a criterion for the triviality of a principal $G$-bundle is 
that its base manifold is contractible to a point (this property can be obtained by pulling back the contractible
loops on the base to the total space, or by using the universal bundle). 

Finally, in the case of the translation bundle $\sbundle{P}{M}{T_4}$, one can chose 
the contractible space $ET_4 = \setR^4$, on which the translations acts freely. Since  $\setR^4$ is contractible to a point, so is the
classifying space $BT_4$, then the bundle $\sbundle{ET_4}{BT_4}{T_4}$ is trivial, and so is 
its isomorphic bundle $\sbundle{P}{M}{T_4}$.

\pagebreak

%--------References
%\bibliography{TEGRNotes.bib}

\end{document}